\documentclass[a4paper]{jpconf}
\usepackage{graphicx}
\usepackage{subfigure}
\begin{document}
\title{Search for $W'\rightarrow t\bar{b}$ in the lepton plus jets final states with the ATLAS detector at the LHC}

\author{Geoffrey GILLES on behalf of the ATLAS collaboration.}

\address{Laboratoire de Physique Corpusculaire de Clermont-Ferrand, \\ Campus des C\'ezeaux - 24 avenue des Landais - 63171 Aubi\`ere Cedex, France.}

\ead{geoffrey.gilles@cern.ch}

\begin{abstract}
This document presents a search for a $W'$ boson, decaying to a top quark and a $b$ quark in an effective coupling approach, using a multivariate method based on boosted decision trees. It reports exclusion limits on the $W'\rightarrow tb$ cross-section times branching ratio and effective couplings as a function of the $W'$-boson mass. The search covers $W'$-boson masses between 0.5 and 3.0 TeV, for right-handed or left-handed $W'$-boson, with 20.3 fb$^{-1}$ of proton-proton collision data produced by the LHC in 2012, at a center-of-mass energy of 8 TeV and collected by the ATLAS detector.
\end{abstract}

\section{Introduction}
Many theoretical approaches beyond the Standard Model involve enhanced symmetries that introduce new charged vector currents carried by new heavy gauge bosons, usually called $W'$. The $W'$ boson can appear as Kaluza-Klein excitations of the $W$ boson in extra-spacetime dimension models \cite{bib1}, as massive right-handed counterpart of the $W$ boson in Left-Right Symmetric models \cite{bib2} in order to restore the parity symmetry at high energy, or in Little Higgs models \cite{bib3} which provide mechanisms to cancel the quadratic divergences that appear in the Higgs-boson mass calculation. In many of these theories, the $W'$ boson is expected to couple more strongly to the 3$^{\rm rd}$ generation of quarks than to the two first ones, opening channels potentially inaccessible to leptonic searches. A search for $W' \rightarrow t\bar{b}$\footnote{In this document we use the notation "$t\bar{b}$" to describe both $W'^+ \rightarrow t\overline{b}$ and $W'^- \rightarrow \overline{t}b$ processes.} signal in an effective coupling approach \cite{bib4} is performed with 20.3 fb$^{-1}$ of proton-proton collisions data produced by the LHC in 2012 and collected by the ATLAS detector \cite{bib5}, at a center-of-mass energy of 8 TeV \cite{bib6}. After defining the event selection criteria, a multivariate technique based on Boosted Decision Trees (BDT) classifiers \cite{bib7}, is used to search for purely right-handed or left handed $W'$ bosons in the mass range of 0.5 -- 3.0 TeV. An additional search for $pp\rightarrow W/W'_L \rightarrow t\bar{b}$ including interference term is also investigated. 

\section{Summary of the search} 
\begin{figure*}[!h!tdp]
\centering
\includegraphics[width=0.35\textwidth]{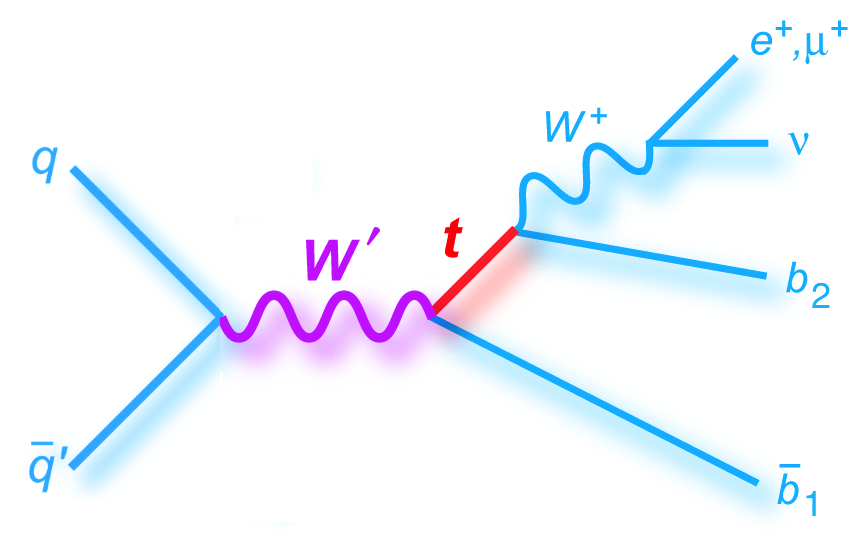}
\caption{Feynman diagram of the $W'$ production and decay chain.}
\label{fig:Signal_Feynman_Diagram}
\end{figure*}

This search is performed in the semi-leptonic final state $W'\rightarrow tb \rightarrow b\overline{b}l\nu$ (Figure \ref{fig:Signal_Feynman_Diagram}), by requiring the presence of a prompt lepton (electron or muon) for a cleaner experimental signature and two $b$-jets in order to reduce the background. 

\begin{figure*}[!h!tdp]
\centering
\subfigure[]{\includegraphics[width=0.34\textwidth]{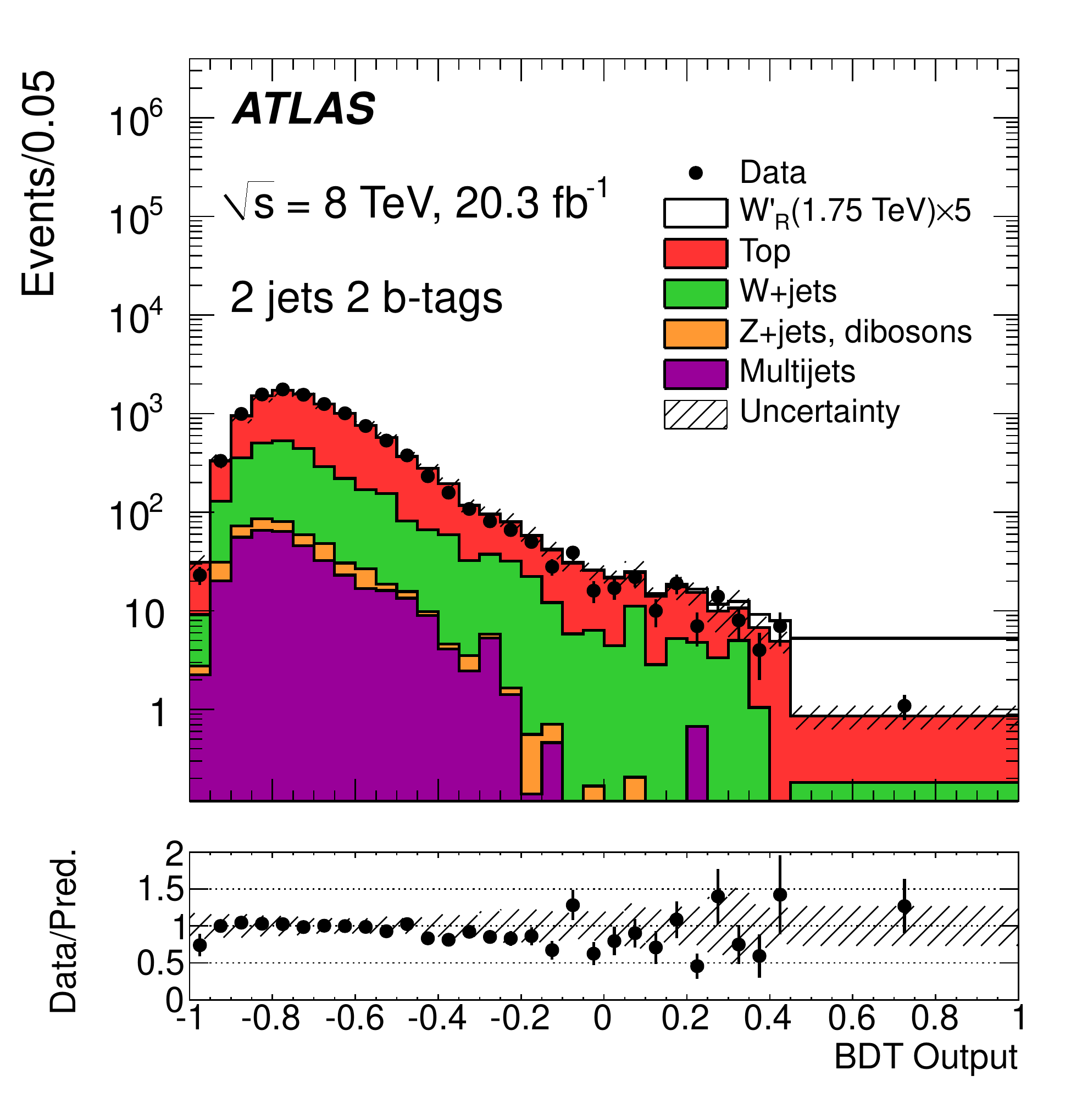}
\label{fig:BDT_SR_2jet}}
\subfigure[]{\includegraphics[width=0.34\textwidth]{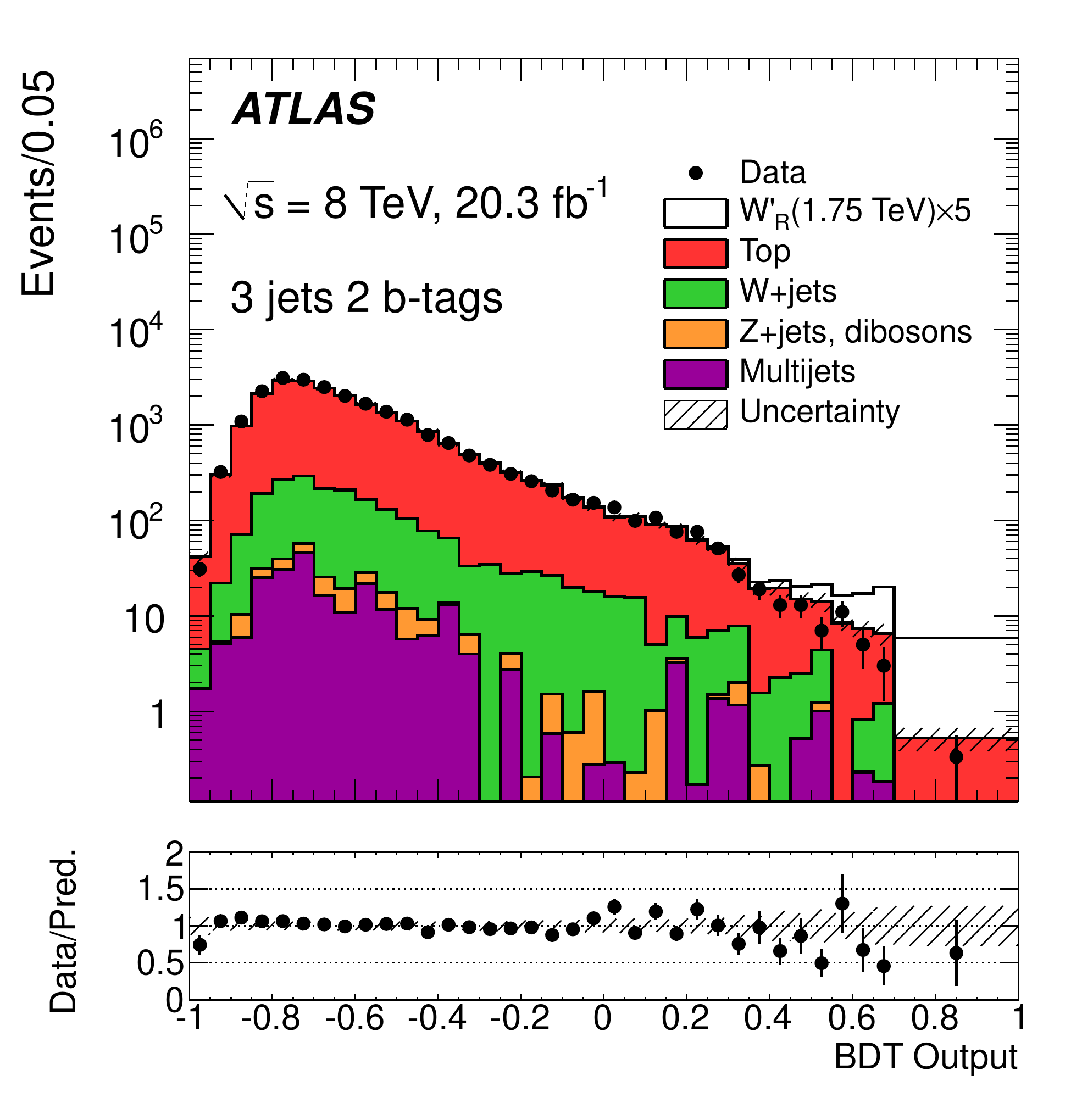}
\label{fig:BDT_SR_3jet}}
\caption{BDT output distributions in the \subref{fig:BDT_SR_2jet} 2-jet and \subref{fig:BDT_SR_3jet} 3-jet signal regions where electron and muon channels are combined. The process labelled "Top" includes $t\bar{t}$ production and all three single top-quark production modes. A signal contribution, amplified by a factor of five, corresponding to a $W'_R$ boson with a mass of 1.75 TeV is shown on top of the background distributions. Uncertainty bands include normalisation uncertainties on all backgrounds and the uncertainty due to the limited size of the simulated samples. 
}
\label{fig:SignalBDT}
\end{figure*}

\noindent In this context, top-quark pairs ($t\overline{t}$) and $W$+jets events are the main backgrounds. Other smaller background processes consist of single top-quark production, diboson and $Z$+jets events. An additional instrumental background stems from multijet production with a jet misidentified as a lepton. The multijet and $W$+jets backgrounds are determined with data-driven methods and all of the remaining background processes are modelled using simulation and scaled to their theoretical predictions. The $W'$ boson is searched for in events with two or three jets, two of which are identified as $b$-jets (2-tag). The signal region used for the search is defined by selecting events where the reconstructed invariant-mass of the $t\bar{b}$ system "$m(t\bar{b})$" is higher than 330 GeV. After the event selection, a BDT algorithm is used to separate a potential $W'$-boson signal from the background. Four BDTs are trained in the 2-jet 2-tag and 3-jet 2-tag channels using right-handed and left-handed $W'$ bosons of mass 1.75 TeV as signal and all of the background processes are weighted to their relative abundance. A set of a dozen of discriminating kinematical variables is used to build the BDTs for each channel where $m(tb)$ and $p_T(t)$ are the two most discriminating variables. Only variables which are well modelled in control regions are selected. Fig \ref{fig:SignalBDT} presents the output distributions of BDTs optimized in the 2-jet and 3-jet signal regions. A signal corresponding to a $W'_R$ boson is shown.

\section{Results}

\begin{figure*}[!h!tdp]
\centering
\subfigure[]{\includegraphics[width=0.32\textwidth]{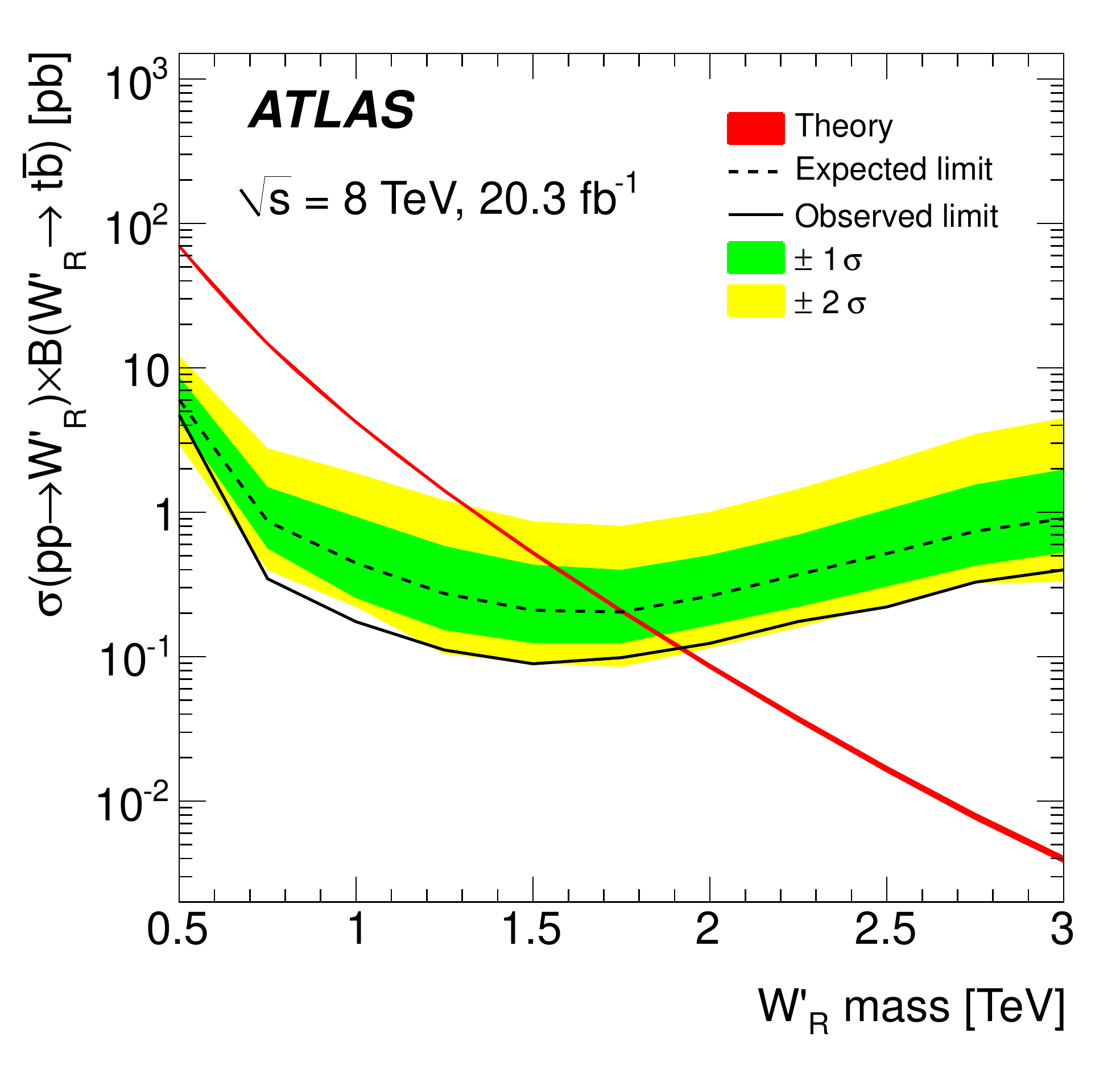}
\label{fig:Limit_right}}
\subfigure[]{\includegraphics[width=0.32\textwidth]{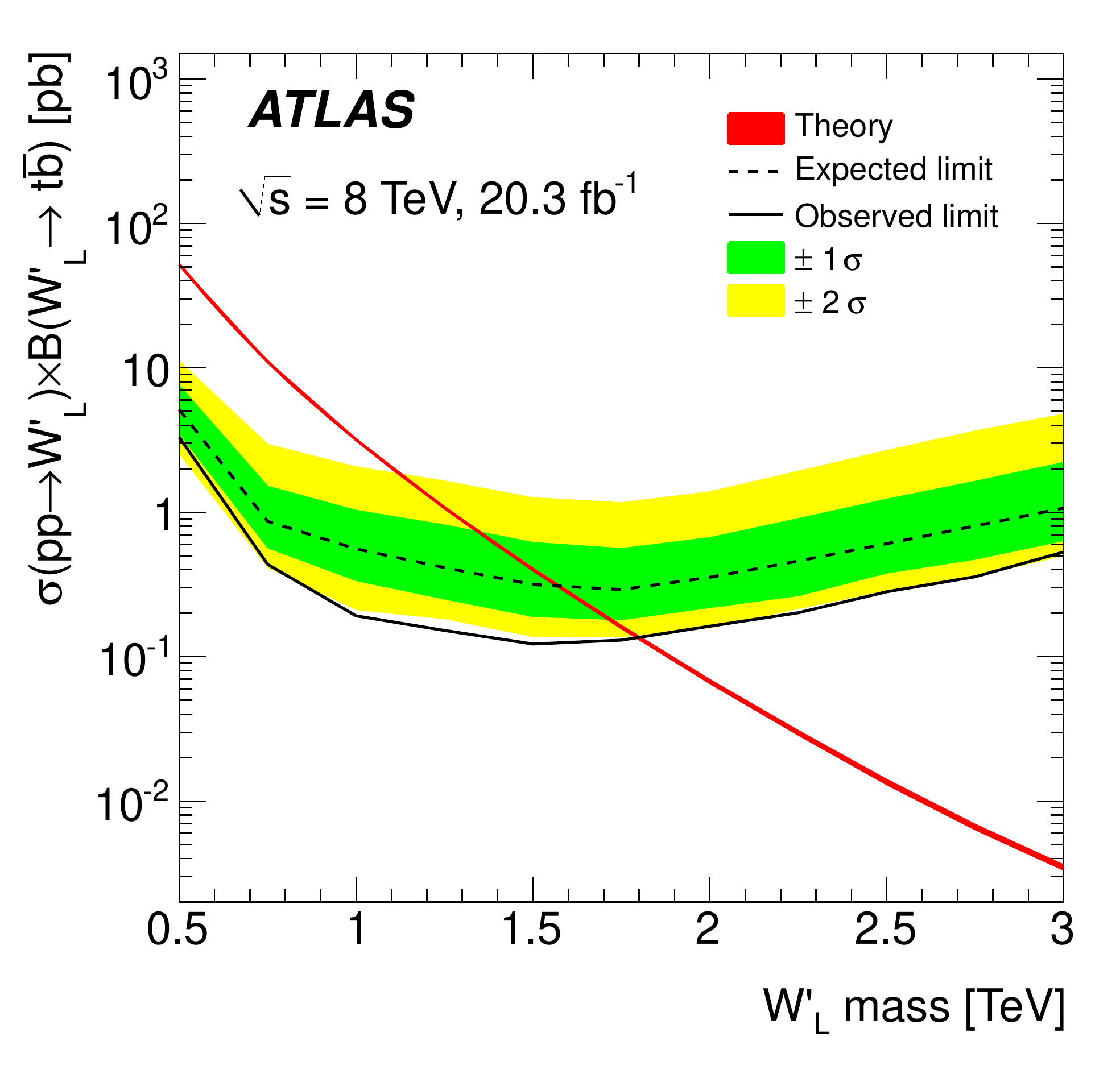}
\label{fig:Limit_left}}
\subfigure[]{\includegraphics[width=0.32\textwidth]{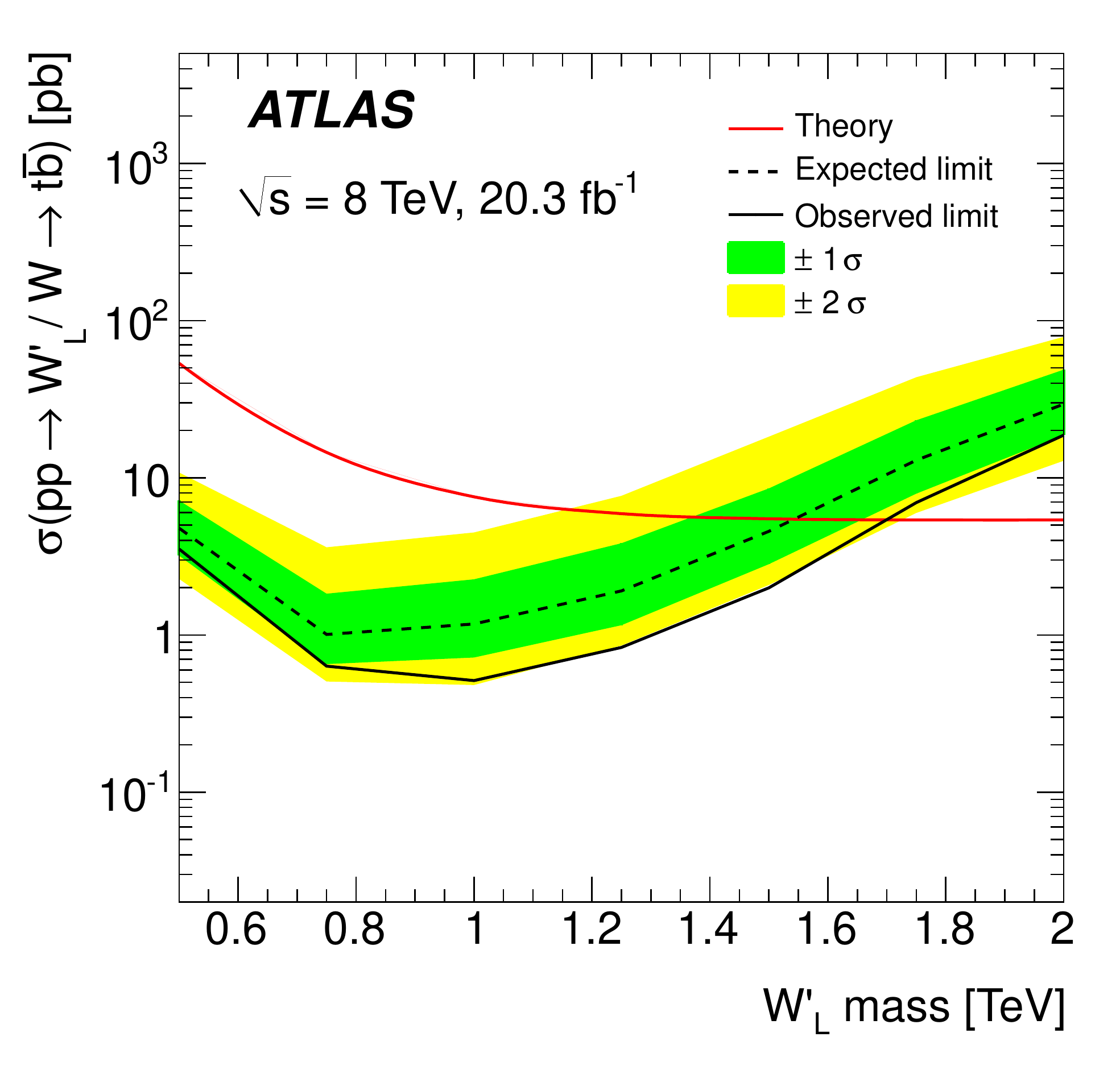}
\label{fig:Limit_left_interf}}
\caption{Observed and expected 95\% CL limits on the $W'$-boson cross-section times branching ratio, as a function of the $W'$-boson mass,
for \subref{fig:Limit_right} right-handed and left-handed $W'$ bosons \subref{fig:Limit_left} with and \subref{fig:Limit_left_interf} without interference. Theoretical predictions of the signal cross-sections [4] are represented by a solid red line. Results include all systematic uncertainties, where uncertainties on $b$-tagging, Monte Carlo generator for $t\bar{t}$ and single-top quark productions and $W$+jets normalization are dominant.
}
\label{fig:LimitBDTXSsyst}
\end{figure*}

No excess of data is observed over the full BDT range. Therefore, the BDT disctributions for 2- and 3-jet channels are combined in a statistical analysis based on the CL$_s$ procedure \cite{bib8}\cite{bib9}, to calculate exclusion limits at 95\% Confidence Level (C.L.) in the production cross-section times branching ratio of the signal as a function of its mass (Fig.\ref{fig:LimitBDTXSsyst}). Results include all systematic uncertainties, where uncertainties on $b$-tagging, Monte Carlo generator for $t\bar{t}$ and single-top quark productions and $W$+jets normalization are dominant. Masses below 1.92 (1.80, 1.70) TeV are excluded for right-handed (left-handed without and with interference) $W'$ bosons.  \\

Limits on the effective coupling ratios $g'/g$ as a function of the $W'$-boson mass are also derived from the limits on the $W'$-boson cross-section.

\begin{figure*}[!h!tdp]
\centering
\subfigure[]{\includegraphics[width=0.32\textwidth]{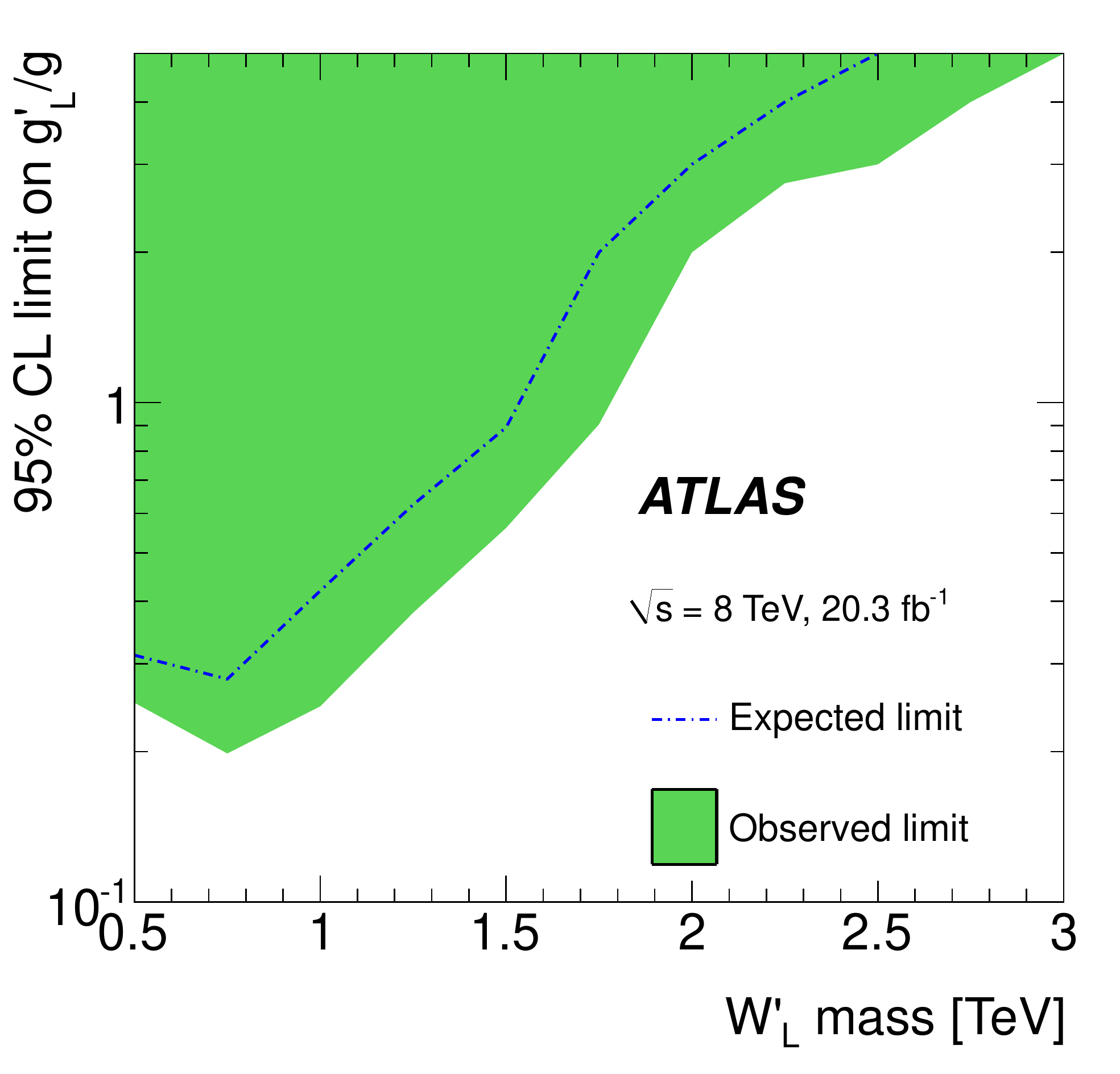}
\label{fig:gprime_left}}
\subfigure[]{\includegraphics[width=0.32\textwidth]{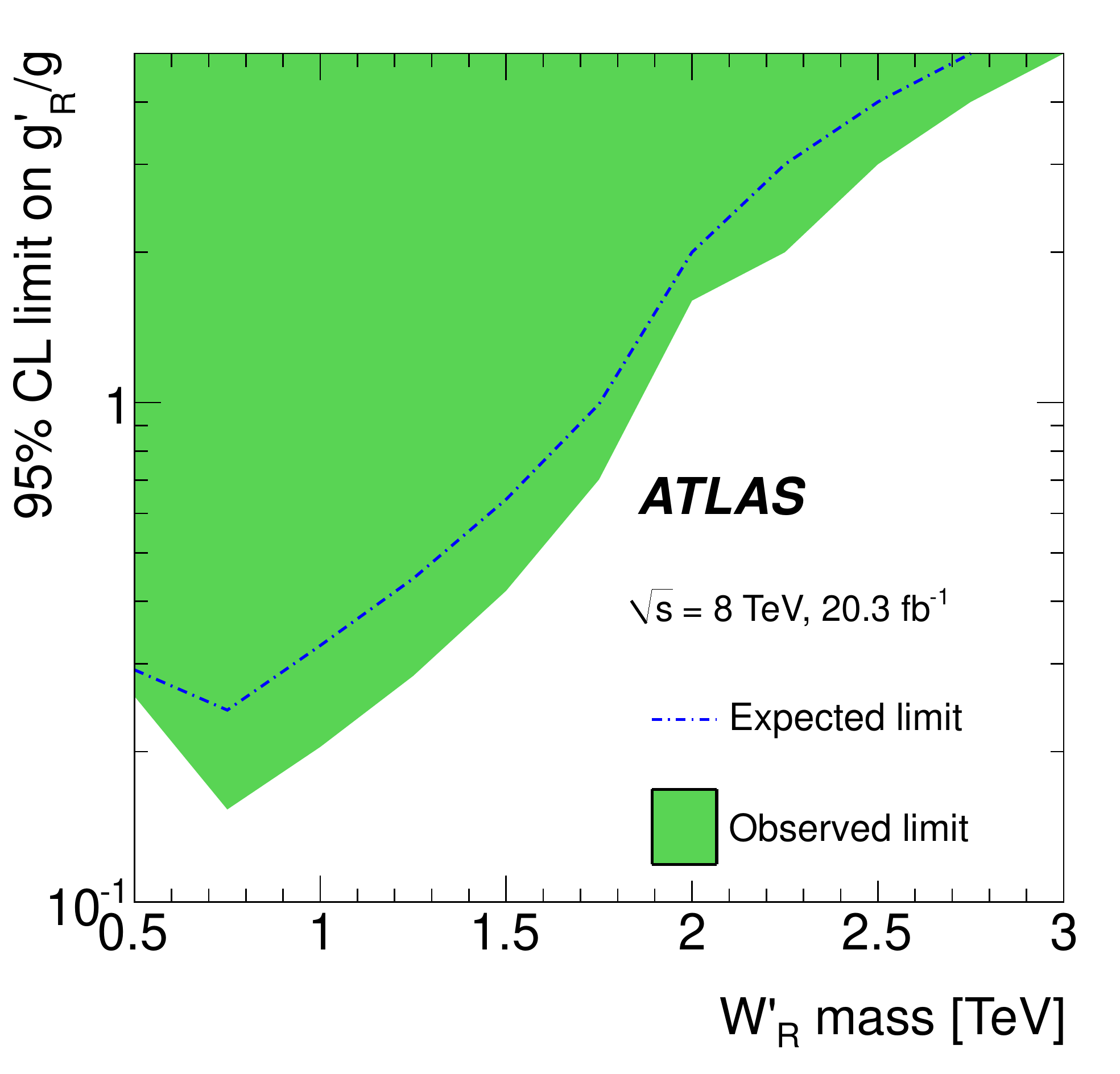}
\label{fig:gprime_right}}
\caption{Observed and expected regions, on the $g'/g$ vs mass of the $W'$-boson plane, that are excluded at 95\% CL,
for~\subref{fig:gprime_left} left-handed (no interference) and~\subref{fig:gprime_right} right-handed $W'$ bosons. 
}
\label{fig:LimitBDTGsyst}
\end{figure*}

\section{Conclusions}

A search for $W'$ bosons, performed with 20.3 fb$^{-1}$ of proton-proton collision data produced by the LHC in 2012, at a center-of-mass energy of 8 TeV  and collected by the ATLAS detector, shows consistency of the data with the Standard Model expectation. No presence of $W'$-boson signal events is observed. Exclusion limits at 95\% C.L. are set on the $W'$-boson mass and on its effective couplings. Masses below 1.92 (1.80, 1.70) TeV are excluded for right-handed (left-handed without and with interference) $W'$ bosons. The lowest observed (expected) limits on $g'/g$, obtained for a $W'$- boson mass of 0.75 TeV, are 0.20 (0.28) and 0.16 (0.24) for left-handed and right-handed W' bosons.

\end{document}